%% file: MAIN_SEIP_format.tex
\documentclass[sigconf]{acmart}

\AtBeginDocument{%
  }

\copyrightyear{2024}
\acmYear{2024}
\setcopyright{acmlicensed}\acmConference[ICSE-SEIP '24]{46th International Conference on Software Engineering: Software Engineering in Practice}{April 14--20, 2024}{Lisbon, Portugal}
\acmBooktitle{46th International Conference on Software Engineering: Software Engineering in Practice (ICSE-SEIP '24), April 14--20, 2024, Lisbon, Portugal}
\acmDOI{10.1145/3639477.3639717}
\acmISBN{979-8-4007-0501-4/24/04}

\acmConference[ICSE 2024]{46th International Conference on Software Engineering}{April 2024}{Lisbon, Portugal}

\input{content/personal_settings}

\begin{document}

\title{\ourtitle}


\author{Doriane Olewicki}
\email{doriane.olewicki@queensu.ca}
\affiliation{%
  \institution{Queen's University}
  \country{Canada}
}
\author{Sarra Habchi}
\email{sarra.habchi@ubisoft.com}
\affiliation{%
  \institution{Ubisoft}
  \country{Canada}
}
\author{Mathieu Nayrolles}
\email{mathieu.nayrolles@ubisoft.com}
\affiliation{%
  \institution{Ubisoft}
  \country{Canada}
}
\author{Mojtaba Faramarzi}
\email{mojtaba.faramarzi@umontreal.ca}
\affiliation{%
  \institution{Université de Montréal}
  \country{Canada}
}
\author{Sarath Chandar}
\email{sarath.chandar@mila.quebec}
\affiliation{%
  \institution{Polytechnique Montréal}
  \country{Canada}
}
\author{Bram Adams}
\email{bram.adams@queensu.ca}
\affiliation{%
  \institution{Queen's University}
  \country{Canada}
}

\renewcommand{\shortauthors}{Olewicki et al.}

\input{content/abstract_keys}

\maketitle

\input{content/body}

\bibliographystyle{IEEEtran}
\balance
\bibliography{acmart.bib}

\end{document}

%% file: content/personal_settings.tex
\usepackage{color,soul}
\usepackage{multirow}
\usepackage{caption}
\usepackage{subcaption}
\usepackage{xspace}
\usepackage{amsmath}
\usepackage{algorithm,algorithmic}
\usepackage{fancybox}
\usepackage{hyperref}
\usepackage{graphics}
\usepackage{tabularx}

\newcommand*{\eg}{{e.g.,}\@\xspace}
\newcommand*{\ie}{{i.e.,}\@\xspace}

\newcommand{\ourtitle}{On the Costs and Benefits of Adopting Lifelong Learning for Software Analytics - Empirical Study on Brown Build\\ and Risk Prediction}
\newcommand{\shorttitle}{On the Costs and Benefits of Adopting Lifelong Learning for Software Analytics}

\newcommand{\RQone}{How does an LL setup perform compared to an RFS setup for software classification tasks?}
\newcommand{\RQoneone}{How beneficial is an NN-based LL compared to state-of-the-practice RFS approaches?}
\newcommand{\RQonetwo}{To what extent does an NN-based LL setup improve the performance of an NN-based RFS setup?}
\newcommand{\RQtwo}{How essential is the use of a memory-based LL setup?}
\newcommand{\RQthree}{How time-efficient is an LL setup compared to an RFS setup?}

%% file: content/abstract_keys.tex
\begin{abstract}

Nowadays, software analytics tools using machine learning (ML) models to, for example, predict the risk of a code change are well established. However, as the goals of a project shift over time, and developers and their habits change, the performance of said models tends to degrade (drift) over time. Current retraining practices typically require retraining a new model from scratch on a large updated dataset when performance decay is observed, thus incurring a computational cost; also there is no continuity between the models as the past model is discarded and ignored during the new model training. Even though the literature has taken interest in online learning approaches, those have rarely been integrated and evaluated in industrial environments.

This paper evaluates the use of lifelong learning (LL) for industrial use cases at Ubisoft, evaluating both the performance and the required computational effort in comparison to the retraining-from-scratch approaches commonly used by the industry. LL is used to continuously build and maintain ML-based software analytics tools using an incremental learner that progressively updates the old model using new data. To avoid so-called ``catastrophic forgetting'' of important older data points, we adopt a replay buffer of older data, which still allows us to drastically reduce the size of the overall training dataset, and hence model training time. 

Empirical evaluation of our LL approach on two industrial use cases, i.e., a brown build detector and a just-in-time risk prediction tool, shows how LL in practice manages to at least match traditional retraining-from-scratch performance in terms of F1-score, while using 3.3-13.7x less data at each update, thus considerably speeding up the model updating process. Considering both the computational effort of updates and the time between model updates, the LL setup needs 2-40x less computational effort than retraining-from-scratch setups, thus clearly showing the potential of LL setups in the industry.
\end{abstract}

\keywords{Software Analytics, Brown build detection, Just-in-Time risk prediction, Lifelong learning, Online learning.}


%% file: content/body.tex
\section{Introduction}
\input{content/intro}

\section{Background and Related Work}
\input{content/relatedWork}



\section{Lifelong Learning for Software analytics}
\input{content/LL_SE}

\section{Empirical study methodology}
\input{content/empirical_metho}

\input{content/results}

\section{Discussion}

\input{content/discussion}

\section{Threats to Validity}
\input{content/threats}

\section{Conclusion}
\input{content/conclusion}

%% file: content/intro.tex
While software analytics models are increasingly used in industrial settings, e.g., for build failure prediction in Continuous Integration (CI)~\cite{saidani2022improving}, test case prioritization~\cite{marijan2022comparative} or code recommendation~\cite{luan2019aroma, schumacher2020improving}, such models 
tend to become outdated over time as they are being used in production~\cite{ekanayake2009tracking}.
MLOps engineers refer to this phenomenon of models and dataset aging as ``concept drift''~\cite{sato2019continuous,gift2021practical,lu2018learning}. The term ``concept'' refers to a segment of a dataset along a timeline having a given data distribution, while ``drift'' refers to the transition between concepts. Concept drift implies that models trained on older segments of data no longer provide information relevant to the prediction in the current segment, potentially harming the performance of the model in production~\cite{gama2014survey}.

As a consequence, in practice, drifting models are discarded and replaced by new models, trained from scratch on the existing datasets for the task.
Such retraining-from-scratch (RFS) setups are expensive to maintain, 
and the training itself becomes more and more computationally expensive as the dataset grows in size~\cite{tsangaratos2016comparison, chen2016xgboost, chung2008minimum}. 
Due to this cost, retraining of models happens sparingly and is usually triggered when a clear performance decay is noticed.
Also, RFS setups ignore prior models' knowledge, which goes against traditional software re-engineering practices~\cite{demeyer2002object}. 

To better deal with concept drift, the domain of machine learning has been exploring online learning approaches. 
These cover various paradigms trying to mimic the human ability to retrain and accumulate knowledge previously learned~\cite{liu2017lifelong}.
In the context of software analytics, prior work incrementally updates the training data with new incoming data, considering resampling approaches to deal with data imbalance~\cite{mcintosh2018fix,tan2015online}, bagging approaches~\cite{cabral2019class,tabassum2020investigation,cabral2022towards} or approaches using memory-based updates~\cite{gao2023keeping}. 
Those works were evaluated on several types of software analytics models in the context of open-source development.

However, such online approaches come with many challenges in practice and have thus rarely been introduced in industrial settings~\cite{barry2023streamai}, which are constrained by efficiency and limited resources for ML training. 
Industries also have expectations regarding ML tools, among which the ability to monitor and validate the model, to have better control over its behaviour.
As such, it is essential to understand both the benefits and costs of online learning approaches in software analytics models in practice, countering 
the drifting nature of software engineering data when training and deploying.

This paper empirically compares the use of online learning approaches and RFS approaches in an industrial setting in terms of performance and computational effort. 
Inspired by prior work on online learning for generative software analytics tasks (e.g., code summarization)~\cite{gao2023keeping}, Natural-Language Processing tasks (e.g., vulnerability detection, and code clone detection)~\cite{gao2023keeping,tan2015online} and other classification tasks (e.g., defect prediction)~\cite{tan2015online, cabral2022towards}, we explore a paradigm of online learning called ``Lifelong learning'' (LL)~\cite{parisi2019continual}.
Our LL setup uses incremental learning models, i.e., Neural Networks (NN), similar to Cabral et al.'s approach~\cite{cabral2019class}, but with the addition of a memory-based mechanism inspired by Gao et al.~\cite{gao2023keeping}.  

Our empirical study focuses on two classification-based software analytics use cases of our industrial partner, Ubisoft, where signs of concept drift have been observed, i.e., brown build detection~\cite{olewicki2022towards} and Just-In-Time (JIT) risk prediction~\cite{nayrolles2018clever}. Brown build detection is related to flaky test analysis, which is one of the emerging topics in software testing in recent years~\cite{olewicki2022towards, bell2018deflaker, alshammari2021flakeflagger}. JIT risk prediction is a generalization of bug prediction models, a staple of software analytics for 20 years, that focuses on the overall risk (e.g., test failure, code complexity) of a change~\cite{rosen2015commit, moser2008comparative}. 
Through empirical evaluation on a total of eight large projects, seven from Ubisoft and one open-source, we address the following research questions: 

\textbf{RQ1: \RQone} 
Comparison with different RFS setups for the two Ubisoft use cases shows that the LL setup significantly outperforms the currently used approaches (+9-31\% F1 score) for 6 out of the 8 studied projects (large effect size), and behaves similarly in the other 2 projects.
We also find that it performs similarly to an NN-based RFS setup in 4/8 cases, significantly better in 3/8 cases and significantly worse in 1/8 case.

\textbf{RQ2: \RQtwo} 
A major issue encountered with LL setups is catastrophic forgetting, where, during updating, the model forgets previously acquired knowledge.  
To address this, we adopt a memory-based approach that trains the model with additional samples of past data points. 
We observe that this approach improves significantly the F1-score compared to a more naive LL setup (+4-26\%) for 3/8 cases, while not showing significant differences in the other 5 cases (-1\% to +6\%).

\textbf{RQ3: \RQthree} 
We compare the computational effort needed to perform LL updates with RFS. Even though the LL setup needs 1.1-8.75 times more updates, the LL updates are computationally less expensive since 3.3-13.7 times less data is necessary for retraining, on average.
The LL setup reports thus 2-40 times less computation effort.

Finally, we discuss trade-offs in terms of the training setups' flexibility to adjust to drift versus continuity in their interpretability.

%% file: content/relatedWork.tex
\subsection{Concept drift}

The concept drift domain studies the evolution of a learning problem on a dataset that is non-stationary~\cite{wang2011concept}. Concept drift in a dataset can be defined as the shift of a concept
~\cite{angluin1988queries}, where the term concept refers to assumptions made on the dataset distributions~\cite{characterizing} in terms of probabilities on the dataset ($X$) and the class labels ($Y$). Usually, the assumptions will involve the prior class probability $P(Y)$, the class conditional probability $P(X|Y)$, or the posterior class probabilities $P(Y|X)$. 

Drift characterization commonly distinguishes sudden, gradual, incremental and recurrent drift~\cite{characterizing}.
In the context of software analytics tools, previous work also suggested approaches to analyze and catalog different types of data drift~\cite{turhan2012dataset,mcintosh2018fix}.
MLOps engineers need to solve the problem of adapting deployed models to concept changes, while having to estimate the right time for such changes to happen. An active approach is recommended to balance the performance evolution of the model with the updating effort~\cite{ditzler2012incremental}.

ML models typically are stationary (``offline''), i.e., once trained and evaluated on a test set (often only using cross-validation~\cite{refaeilzadeh2009cross}), they are deployed in production as-is. However, the real world encountered in production is dynamic and continuously changing, and the occurrence of drift gradually decreases the performance of the deployed models. To avoid the costs and loss of model continuity of RFS update approaches, researchers have identified the need for developing dynamic/online models~\cite{brenowitz2020machine}.

\subsection{Online Learning for software analytics}

In the context of software analytics, previous works have studied several aspects of online learning leveraging open-source datasets~\cite{cabral2019class, tan2015online, gao2023keeping} and closed-source datasets covering periods of less than a year~\cite{tabassum2020investigation}. 
Tan et al.~\cite{tan2015online} proposed an online learning approach that continuously updates the training dataset with new data points in the context of defect prediction, then uses resampling techniques and updatable learning algorithms based on recomputed probabilities (Naive Bayes) or clustering.
Gao et Al. explored online learning for pre-trained NLP models in the context of code summarization, vulnerability detection and code clone detection with a memory-based approach (or replay buffer)~\cite{gao2023keeping}.

Other works for online learning rely on bagging approaches, where bagged incremental models are retrained in two cumulative ways: (1) over time with every newly labeled data point and (2) with regular ``replay'' trainings involving past data points selected using resampling (to deal with imbalanced data)~\cite{cabral2019class,cabral2022towards,tabassum2020investigation}. 
Although the results show that bagging tends to improve performance through regularization~\cite{goodfellow2013empirical}, bagging approaches multiply the computational effort by the number of NN models in the bag, thus increasing considerably the computational effort, especially for the replay trainings, which require more data. 
Our goal is to have a lightweight approach allowing frequent incremental training using limited resources, which is why we consider a single model trained in an LL setup. 
Secondly, updating the model at each new data point is not practical as this requires substantially more effort for storing the model and for the MLOps engineers to validate the model.

Industry has high expectations regarding the use of ML tools, among which the ability to monitor and validate a model before its deployment in production and the ability to roll back to stable versions when needed~\cite{barry2023streamai}.
With models constantly evolving on each new labeled data input, monitoring becomes challenging and the validation of each new model update by MLOps engineers becomes unfeasible, making the adoption of software analytics tools risky for tasks related to critical parts of project development. 
Then, to be able to perform a roll-back when needed, each model update should back up a copy of the previous model version's weights, increasing considerably the storage complexity of that type of online setups. 
We thus need to find a trade-off between having regular updates and 
still allowing human monitoring of the updated models.

Inspired by the above works, our approach uses a single 
incremental learner~\cite{tan2015online,gao2023keeping} in the form of an NN model~\cite{cabral2022towards} that is incrementally updated with both new data points and a replay buffer of older data points~\cite{gao2023keeping}. 
We evaluate the approach on closed-source datasets spanning up to 9 years.
Furthermore, as past work has not evaluated the computational effort of online learning, we evaluate computation time of retraining in relation to update frequency.  

\input{tables/algo}

\subsection{Lifelong Learning}
\label{LL_background}

Lifelong learning, also known as continual learning, is a paradigm of online learning that dynamically evolves an existing model based on a stream of information, similar to how humans and animals acquire knowledge, maintain it, and transfer it to new situations~\cite{parisi2019continual}.
Computational systems and agents need to be able to adapt their knowledge over long time periods and multiple tasks, especially when data distributions shift dynamically over time. 
A traditional LL setup adapts the model over time as new data (e.g., new tasks) are made available.

The main challenge with lifelong learning is so-called ``catastrophic forgetting'', which occurs when training a neural network with new data interferes with, or even completely overwrites, previously acquired knowledge~\cite{goodfellow2013empirical}. This leads to a drop in model performance, especially for previously learned tasks for which the knowledge is not available anymore~\cite{kemker2018measuring}.
The learning system must thus find a balance between retraining past knowledge and molding knowledge to new information to still be able to learn new tasks~\cite{ditzler2015learning}, while scaling to online learning.

The oldest method in use to reduce catastrophic forgetting is using memory.
A memory, also called replay buffer, is built over time from data samples and is regularly replayed during retraining, in addition to new samples, to remind the model of past observations~\cite{isele2018selective}. 
Since keeping the entire historical data stream quickly becomes unmanageable in terms of both data storage and training time, and is also counterproductive due to outdated data, usually a simpler approach is to use a First-In-First-Out (FIFO) queue to limit the memory size~\cite{riemer2019scalable}.
Selecting the replay buffer samples to cover a larger diversity can significantly improve the performance of LL~\cite{lanka2018archer, eysenbach2019search}, similar to approaches like dynamic NN size or regularization~\cite{parisi2017lifelong,benna2016computational, kirkpatrick2017overcoming}.

%% file: tables/algo.tex
\begin{algorithm}[t]
\small
\caption{Pseudo-code for training of an NN.}\label{algo_nn}
\begin{algorithmic}
\REQUIRE{trainset with dataset "X" and labels "y", NN, epoch}
\FOR{iter in epoch}
 \STATE Split trainset in minibatches
 \FOR{minibatch in trainset}
 \STATE pred = NN(minibatch[X]) //model's outputs on mini-batch
 \STATE loss = lossfun(minibatch[y], pred) //loss versus labeled data
 \STATE net.update(optimizer(NN, loss)) //update model weights
 \ENDFOR
 \ENDFOR
\end{algorithmic}
\end{algorithm}

%% file: content/LL_SE.tex
\subsection{Multi-layer perceptron neural networks} \label{sec:MLP}

An LL setup requires an incremental learning algorithm, such as multi-layer perceptron (MLP) neural networks. 
A basic MLP has $L$ layers of size $N$. Between each pair of neighboring hidden layers, an activation function (\eg ReLU) is applied to the output of one layer before sending it to the next layer in order to have a non-linear transformation~\cite{gardner1998artificial}. After the final layer, a sigmoid activation function is applied to bound the final output values between 0 and 1 to obtain probabilities as output. Algorithm~\ref{algo_nn} shows the pseudo-code for a typical MLP training process~\cite{gardner1998artificial}. In essence, in each iteration (``epoch'') the training algorithm tries to minimize with an optimizer the ``loss'' between the current model's prediction on subsets (``mini-batch'') of the training set and the expected output. The complexity of this algorithm is linearly dependent on the number of epochs and the size of the training set (divided by the size of a mini-batch).

Implementation-wise, the loss function we use is Binary Cross Entropy (BCE), while the optimizer we consider is Adam (with hyperparameter learning rate $lr$). 
Both were chosen after preliminary evaluation instead of, respectively, the loss functions Mean Squared Error (MSE) and Mean Absolute Error (MAE), and the optimizer Stochastic Gradient Descent (SGD).
The MLP training is done using mini-batches~\cite{li2014efficient} of size 20 and with 50 epochs at each training, since our case studies showed that this was enough to pass the under-training stage and to reach the over-training phase. Once all 50 epochs are run, the model stage at the best epoch (in terms of performance on the validation set) is kept, ready for production. 
The incremental nature of this algorithm refers to its capacity of updating the pretrained weights of an existing NN model with an updated training set.

\subsection{Data splitting} 

In an actual production setup, the ML-models' datasets are split chronologically such that the training, validation, and test sets do not overlap. 
In the case of LL, these sets have to keep up with time, as the goal of LL is to update models based on recently encountered data.
For this reason, in an LL setup, the dataset is split into successive ``time groups''. 
Such groups are split according to the number of consecutive commits encountered, the group size ($GS$) being a hyperparameter of the setup. 
We opted to split based on number of commits instead of a fixed time period, since software engineering datasets have varying activity over time with some weeks having little to no data. 
Furthermore, in an LL setup, each model update needs to train on time groups with a decent amount of data in order not to overfit on new data and not to unlearn underrepresented classes' predictions because of sparse training sets.

\subsection{Lifelong learning training process}

\begin{figure}[t]
	\centering
	\includegraphics[width=.45\textwidth]{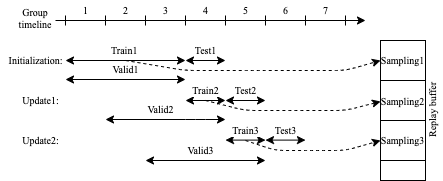}
	\caption{LL setup\label{fig:LL_setup} with validation set, initial training set and replay buffer of size $VWin=ITWin=RBWin=3$, resp.
	}
\end{figure}

\subsubsection{Overall algorithm}

The LL setup has two different types of time steps: one initialization step, followed by multiple update time steps. 
At each time step of the LL training process, the algorithm shifts up one time group of data ($GS$), then defines separate sub-datasets for the training, validation, and testing of the NN. Figure~\ref{fig:LL_setup} shows an example of the first three time steps (named ``Initialization'', ``Update1'' and ``Update2'') of an LL setup with a sliding window validation set of size $VWin=3$ (as an example).

A given time step's three sub-datasets never overlap amongst each other; we also avoid certain kinds of overlap across time steps. In particular, every data point that ever appeared in a validation set will never appear in a training set and vice-versa. This is achieved through an initial random sampling while respecting the class representations in each set. Note that the test set of a given time step by definition will overlap with the next step's training set. To illustrate, the data of Train1 will never overlap with Valid1 or Valid2, while the data of Valid1 and Valid2 can overlap. 

The validation sets are created with data points of a sliding window of size $VWin$, which is different from traditional LL setups where the whole timeline would be covered. This choice was made because we suspect projects to exhibit concept drift, hence progressively filtering out older time steps allows the model to eventually forget outdated knowledge.

However, in order to limit the catastrophic forgetting of the model, we define a replay buffer. 
This replay buffer is built as we go through time steps, by randomly selecting a portion of new training data points found since the last update, \ie sampled from group $t$.
That sample is balanced, \ie the same number of data points is selected of each classification label, to have a significant representation of both labels at training time. This does not impact the evaluation as the validation and test set have the original data distribution.
As mentioned in Section~\ref{LL_background}, we need to keep the replay buffer's size limited in order to keep incremental retraining time manageable and to avoid old data from harming the model~\cite{olewicki2022towards}. 
Hence, to ignore older data points, we use a First-In-First-Out (FIFO) approach with window size ($RBWin$) as hyperparameter.

\subsubsection{Time step 1 (Initialization)}

During the initialization step, the MLP is created and trained. We select an initial training window of size $ITWin$ groups, whose hyperparameters are tuned on a validation set of $VWin$ groups. Then, the trained model is deployed to predict each data point in the next group ($ITWin+1$). 
During the initialization step, we can define the different sets as: 
	\begin{itemize}
		\item Replay buffer: buffer = \{\}
		\item Training set: 
		train = \{ points $d$ sampled from group $i~|~i\in [1; ITWin]$ AND $d  \notin valid$ \} ;
		\item Validation set: valid =
		\{ points $d$ sampled from group $i~|~i\in]ITWin - VWin; ITWin]$ AND $d  \notin train$ \};
		\item Test set: test = \{ points $d$ in group $i~|~i=ITWin+1$ \}
	\end{itemize}

After training, balanced random sampling is applied to the training set, and the sampled data is added to the replay buffer. 
	
\subsubsection{Subsequent time steps (Update)}
	Once the initial model has been deployed, the strength of LL starts to show up on subsequent time steps, when the MLP model will be updated incrementally. At a given time step $t$, 
	the deployed model from $t-1$ is updated with the new data of time group $t$ that has been labeled
 , as well as the replay buffer's current state, containing samples anterior to $t$. At the start of time step $t$, we can define the different sets as:
	\begin{itemize}
		\item New data set: new\_data = \{ points $d$ sampled in group $i~|~i=t$ AND $d \notin valid$ \};
		\item Replay buffer: buffer = \{ points $d$ sampled from group $i~|~i\in[t - RBWin; t[$ \};
		\item Training set: train = New data set $\cup$ Replay buffer;
		\item Validation set: valid = \{ points $d$ sampled from group $i~|~i\in]t - VWin; t]$ AND $d \notin train$\} ;
		\item Test set: test = \{ points $d$ in group $i~|~i=t+1$\}.
	\end{itemize}
Again, the test set is not available at the time of retraining, but represents data encountered after deployment. 
Furthermore, the FIFO policy for the replay buffer is incorporated above by removing all data points from before group $t-RBWin$. At the end of the training phase, a balanced sampling is done on data group $t$ and added to the replay buffer.

%% file: content/empirical_metho.tex
\subsection{Use cases}

Our empirical study validates LL on two software analytics tools used in the CI pipelines of our industrial partner, Ubisoft, with historic data from 7 closed-source projects that each have been using one of the software analytics tools for 1.5 to 9 years. We also consider one open-source project that was used as a reference dataset when developing the brown build detection tool at Ubisoft, as the data distribution is similar to the closed-source ones. 
One tool performs brown build detection~\cite{olewicki2022towards}, while the other performs JIT risk prediction~\cite{nayrolles2018clever}. The features used in these tools are derived from build logs and version control metadata, respectively. 
Ubisoft has observed concept drift on those datasets through performance decay, as the studied projects, all related to game development, follow development steps/gates spanning across the different development milestones over time. The changing nature of development activities along these gates motivates the need to update software analytics tools regularly.

\input{tables/data_brown}

\subsubsection{Brown Build detection}

\ \\{\noindent}\textbf{Description.} CI evaluates the quality and risk of new code changes submitted to the integration pipeline by executing builds and test jobs~\cite{shahin2017continuous}.
In theory, successful builds validate the changes, whereas failed builds indicate the need to rectify a code change before integrating it to the project. In practice, the difference between successful and failed builds is blurred by so-called ``brown builds''.

A brown build (or flaky build) is defined as an intermittent build job failure that would change to success just by rerunning~\cite{lampel2021life}.
Those inconsistent build failures are often attributed to the build scripts themselves~\cite{ghaleb2019studying,gallaba2018noise} (\eg missing dependencies and network access) or to flaky tests~\cite{dong2020concurrency,bell2018deflaker,herzig2015empirically}.
Our industrial partner observed that brown builds make developers lose trust in the CI system. Whenever a build fails, instead of verifying if the code is actually faulty, they would force a manual rerun of the build just to see if the result changes. This unnecessarily increases the build activity on the integration pipeline and the integration time. 

Previous approaches to predict whether a build is brown or safe are based on flaky test detection approaches~\cite{bell2018deflaker,alshammari2021flakeflagger,Haben2021Replication} or based on build log vocabulary extraction~\cite{lampel2021life,olewicki2022towards}. Ubisoft developed a latter type of model.
This approach takes as input build logs, processes them to extract the best (series of) words using KBest feature selection, then vectorizes each trace with TF-IDF. The vectors are then used as input for an XGBoost-based model. 

{\noindent}\textbf{Data extraction.} Table~\ref{info_proj} describes the projects for brown build detection used in this paper. 
Our study focuses on three closed-source projects provided by our industrial partner and one open-source project (Brown\_OSS).
The Brown Failure Rate $\beta$ refers to the percentage of brown job failures across all job failures, including reruns of builds. The studied projects were chosen to have at least 1 year of data with a consistent build job activity over time (at least 1k jobs per week). The projects differ in languages, number of jobs, and in $\beta$. Project Brown\_OSS corresponds to the Graphviz~\cite{graphiz} project, a graph visualization project publicly available on GitLab. We include this project in our study since it has a non-trivial brown failure ratio of 13\% that falls in between the other projects' brown failure rate.

For the four projects, the dataset extracted is composed of build logs and CI metadata such as the job's name, number, associated commit id, and timestamp information. This data is extracted from GitLab~\cite{gitlab}. We also compute additional features about the number of past reruns of the job and the number of commits since the last brown job. These features are used to include information about the CI job's context and the moment it was run. 

To reduce information contamination among build jobs from reruns, all the jobs allocated to a given commit and build job name appear in the same subset. This implies that despite the choice for a fixed number of \textit{(commits,job name)} per group, the size of the groups may vary depending on how many reruns were triggered.

\subsubsection{Just-in-time risk prediction}

\input{tables/risk_feat}
\input{tables/train_setup}

\ \\{\noindent}\textbf{Description.} JIT risk prediction predicts the riskiness of a code change based on metrics related to the change, such as the number of lines added/ modified/ removed, developer's experience, etc.~\cite{kamei2012large}. The term ``just-in-time'' refers to the fact that the detection is done while reviewing the code change instead of right before releasing. This has the advantage that the code change is fresh in the developers' mind.
Research has focused on making the prediction explainable, offering mitigation techniques to the developers to improve their code change~\cite{shihab2012industrial}. Furthermore, researchers have explored both 
tree-based models~\cite{nayrolles2018clever} and neural network models~\cite{qiao2019effort}.

In Ubisoft's implementation of this approach, features are fed to an XGBoost-based model~\cite{chen2015xgboost}.
The tool is currently used in almost all the projects of our industrial partner as a predictor for developers on the riskiness of their code changes or, in some cases, as an integration step.

{\noindent}\textbf{Data extraction.} We study 4 different closed-source projects of our industrial partner, see Table~\ref{info_proj}.
A code change is labeled as risky if it is tagged as bug-introducing in a Jira ticket.
In principle, Jira uses the SZZ approach~\cite{da2016framework} to derive bug introducing changes from bug fix data, but labels can also be manually overridden by developers, e.g., if they deem an SZZ blame to be incorrect.

The features used for the risk prediction, described in Table~\ref{risk:info_feat}, involve metadata about code changes. 
The ``approvers'' are reviewers that can decide which code changes can be integrated into an upcoming release. All code changes are reviewed by at least one reviewer. When nearing a release date, the code changes are approved by an approver to be integrated in the upcoming release (or pushed to the next release).

\subsection{Baselines}

The metric used to evaluate and optimize the performance of the approaches for both use cases is F1-score ($F1$=$\frac{2}{Pre^{-1} + Rec^{-1}}$). The F1-score is the harmonic mean of the precision ($Pre$=$\frac{TP}{TP + FP}$) and recall ($Rec$=$\frac{TP}{TP + FN}$).
We also report the G-Mean ($GM$=$\sqrt{Rec \times Rec_{safe}}$), with $Rec_{safe}$=$\frac{TN}{TN+FP}$, also called specificity. We chose to use the latter two metrics because of the unbalanced nature of our datasets in terms of $\beta$ and $RR$ in Tables~\ref{info_proj}.

As baseline, we use the 8 setups of Table~\ref{tab:trainingsetup}. For both use cases, there is a naive AlwaysBrown/Risky baseline corresponding to always predicting builds/commits to be brown/risky. Those naive baselines are used as context regarding the expected performance in comparison to $\beta$ and $RR$. For brown build analysis, we also include 2 RFS baselines proposed by Olewicki et al.~\cite{olewicki2022towards}, respectively RFS with weekly scheduling (Weekly XGB) or scheduling based on model performance decay (Heuri XGB).
For JIT risk prediction, the ML architecture was not revealed to us, apart from learning that it is XGBoost-based and retrained from scratch every 6-8 weeks, using the full data history. Nevertheless, we did receive the historical prediction results of the actual model (RealPred), a highly relevant baseline consisting of the performance of the real (RFS) model. 

Finally, for both use cases, we include a baseline retraining setup for MLP models using RFS (RFS MLP). This baseline is representative of state-of-the-practice RFS setups and setups relying on the evolutive sampling of the training set~\cite{liu2017lifelong}.
We schedule the retraining of the RFS MLP baseline at every group size, similar to how the LL retraining of our approach is scheduled.


\input{tables/RQ1_res}

%% file: tables/data_brown.tex
\begin{table}[t]
	\centering
	\caption{\label{info_proj} Studied projects for Brown build and JIT risk prediction. The Brown failure rate represents the percentage of brown jobs among the failed build jobs and the risky rate the percentage of risky commits. }

	\begin{tabularx}{.47\textwidth}{|c|X|X|X|X|}
		\hline
		\multirow{3}{*}{\textbf{Project}}&\textbf{Period [month]}&\textbf{\#jobs}&\textbf{\#brown- jobs}&\textbf{Brown failure}\\
		&&&&\textbf{ rate ($\beta$)}\\
		\hline
		Brown\_1&18&23k&3k&30\%\\
		Brown\_2&29&63k&8k&37\%\\
		Brown\_3&17&22k&6k&5\%\\
		Brown\_OSS&43&47k&1k&13\%\\
		\hline
  \hline
		\multirow{2}{*}{\textbf{Projects}}&\textbf{Period [years]}&\textbf{\#commits}&\textbf{\#risky- commits}&\textbf{Risky rate \textbf{($RR$)}}\\
		\hline
		Risk\_1&5&50k&12k&24\%\\
		Risk\_2&4&49k&13k&27\%\\
		Risk\_3&9&82k&20k&24\%\\
		Risk\_4&5&57k&9k&15\%\\
		\hline
	\end{tabularx}

\end{table}

%% file: tables/risk_feat.tex
\begin{table}[t]
	\centering
\caption{\label{risk:info_feat} Features used for Just-In-Time risk prediction.}
	
	\begin{tabularx}{.47\textwidth}{|p{2cm}|X|}
		\hline
		\textbf{Name}&\textbf{Description}\\
		\hline
		Subsystems&\#Modified subsystems. A subsystem is a directory containing code files.\\
		Files& \#Modified, added or deleted files.\\
		Lines added& \#Added lines of code.\\
		Lines deleted& \#Deleted lines of code.\\
		Lines total& \#Modified lines of code (added and deleted).\\
		Devs& \#Developers who changed the modified files.\\
		Age& Average age of the modified files.\\
		Unique changes& \#Unique changes.\\
		Experience& Experience of the developer who submitted the code change. All types of experiences are measured in terms of contributions weighted over time $\frac{1}{(m-1)}$ where $m$ is the \#months since the contribution.  \\
		Reviewers& \#Reviewers assigned to the code change.\\
		Reviewer experience& \#commits the reviewer has reviewed regarding the targeted files. \\
		Reviewer dev. experience& \#commits the reviewer has developed regarding the targeted files.\\
		Reviewer subsystem exper.& \#commits the reviewer has reviewed regarding the targeted files' subsystems.\\
		Approvers& \#Approvers assigned to the code change.\\
		Approver experience& \#commits the approver has reviewed regarding the targeted files.\\
		Approver dev. experience&\#commits the approver has developed regarding the targeted files.\\
		Approver subsystem exper.& \#commits the approver has reviewed regarding the targeted files' subsystems.\\
		FilesRiskiness & Historical ratio of bug introducing changes in the modified files.\\
		Subsystems-Riskiness & Historical ratio of bug introducing changes in the modified subsystems.\\
		Daystorelease&\#Days to the release date of next update of the software product.\\
		\hline
	\end{tabularx}
\end{table}

%% file: tables/train_setup.tex
\begin{table*}[t]
    \centering
    \caption{Training setup for LL and baselines.}
    \label{tab:trainingsetup}
    \begin{tabular}{|c|c|c|p{11.5cm}|}
    \hline
         &Training setup &Use case& Description  \\
         \hline
         \multirow{8}{*}{Baselines}
         &Weekly XGB&Brown& RFS setup on XGBoost model with training window retrained every week~\cite{olewicki2022towards} (training set $\sim$.5-1k jobs).\\
         &Heuri XGB&Brown&RFS setup on XGBoost model with a training schedule based on performance decay \cite{olewicki2022towards} (training set $\sim$.5-1k jobs).\\
         &Real Pred&Risk&Industrial collaborator-provided predictions of RFS setup on XGBoost model (retrained every 6-8 weeks, training set$\sim$all data available).\\
         &RFS MLP&Brown \& Risk&RFS setup on MLP with all the available labeled data from the start of the dataset to the training point, retrained on a group-based schedule (every $GS$).\\
         \hline
         \multirow{2}{*}{LL setups}&LL noRB&Brown \& Risk& LL setup on MLP, retrained on a group-based schedule (every $GS$) without replay buffer.\\
         &LL&Brown \& Risk&LL setup on MLP, retrained on a group-based schedule (every $GS$) with replay buffer.\\
         \hline
    \end{tabular}
\end{table*}

%% file: tables/RQ1_res.tex
\begin{table*}[t]
    \centering
    \caption{Performance after hyper-parameter optimization for the LL setups and baselines. We report the average test results out of 5 runs. The \textcolor{cyan}{\textbf{bold blue}} values indicate the best F1-score (F1) and G-mean (GM) per project.
    For each project, Kruskal-Wallis tests ($\alpha=.05$) between the F1-score/G-mean of runs of the LL setup and of the other setups are performed. If significant, the Cliff's delta effect size is shown as ``+'' (large effect), otherwise ``/'' is shown.}
    \label{CL_hyper_res}
    \begin{tabularx}{\textwidth}{|l|XXXXX|XXXXX|XXXXX|XXXXX|}
    	\hline
    	\multirow{3}{*}{Model}
            & \multicolumn{5}{c|}{Project Brown\_1} 
    	& \multicolumn{5}{c|}{Project Brown\_2}
            & \multicolumn{5}{c|}{Project Brown\_3} 
    	& \multicolumn{5}{c|}{Project Brown\_OSS} \\ \cline{2-21}
    	&\multicolumn{1}{l|}{\multirow{2}{*}{F1}}&\multicolumn{1}{l|}{\multirow{2}{*}{GM}}&\multicolumn{2}{c|}{Brown}&\multicolumn{1}{p{.2cm}|}{Safe}
    	&\multicolumn{1}{l|}{\multirow{2}{*}{F1}}&\multicolumn{1}{l|}{\multirow{2}{*}{GM}}&\multicolumn{2}{c|}{Brown}&\multicolumn{1}{p{.2cm}|}{Safe}
    	&\multicolumn{1}{l|}{\multirow{2}{*}{F1}}&\multicolumn{1}{l|}{\multirow{2}{*}{GM}}&\multicolumn{2}{c|}{Brown}&\multicolumn{1}{p{.2cm}|}{Safe}
    	&\multicolumn{1}{l|}{\multirow{2}{*}{F1}}&\multicolumn{1}{l|}{\multirow{2}{*}{GM}}&\multicolumn{2}{c|}{Brown}&\multicolumn{1}{p{.2cm}|}{Safe}
    	\\
    	&\multicolumn{1}{p{.2cm}|}{}&\multicolumn{1}{p{.2cm}|}{}&Pre&\multicolumn{1}{p{.2cm}|}{Rec}&\multicolumn{1}{p{.2cm}|}{Rec}
    	&\multicolumn{1}{p{.2cm}|}{}&\multicolumn{1}{p{.2cm}|}{}&Pre&\multicolumn{1}{p{.2cm}|}{Rec}&\multicolumn{1}{p{.2cm}|}{Rec}
    	&\multicolumn{1}{p{.2cm}|}{}&\multicolumn{1}{p{.2cm}|}{}&Pre&\multicolumn{1}{p{.2cm}|}{Rec}&\multicolumn{1}{p{.2cm}|}{Rec}
    	&\multicolumn{1}{p{.2cm}|}{}&\multicolumn{1}{p{.2cm}|}{}&Pre&\multicolumn{1}{p{.2cm}|}{Rec}&\multicolumn{1}{p{.2cm}|}{Rec}\\ 
    	\hline
        Weekly XGB&56~+&66~+&59&54&81
    	&67~/&69~+&75&61&79
    	&22~+ &37~+&45&15&95
    	&44~+&66~+&44&45&98\\
    	Heuri XGB&60~+&71~+&61&60&83
    	&67~/&70~+&73&62&79
    	&27~+&43~+&50&19&96
    	&44~+&66~+&44&45&98\\
    	RFS MLP&
    	64~+ & 69~/&53&80&59&
    	70~/&70~/&68&72&69&
    	\textcolor{cyan}{\textbf{36}}~+&\textcolor{cyan}{\textbf{62}}~+&28&51&75&
    	28~+  &  68~+  &  19   &  55   &  84\\
     \hline
    	LL noRB
    	&  66~/  &  70~/  &  54   &  83   &  58
       &  69~/  &  67~/  &  64   &  74   &  62
        &  30~/  &  57~/  &  21   &  52   &  63
    &  49~+  &  81~+  &  33   &  100  &  65 \\
		LL&\textcolor{cyan}{\textbf{72}}&\textcolor{cyan}{\textbf{77}} &{64}&{83}&{72}
	&\textcolor{cyan}{\textbf{71}}&\textcolor{cyan}{\textbf{71}} &{68}&{74}&{68}
	    & 31 &57&{23}&{47}&{69}
    	&\textcolor{cyan}{\textbf{75}}&\textcolor{cyan}{\textbf{94}} &{60}&{100}&{89}\\
    	
    	\hline
    	\hline
            \multirow{3}{*}{Model}
    	& \multicolumn{5}{c|}{Project Risk\_1} 
    	& \multicolumn{5}{c|}{Project Risk\_2}
            & \multicolumn{5}{c|}{Project Risk\_3} 
    	& \multicolumn{5}{c|}{Project Risk\_4} \\ \cline{2-21}
    	&\multicolumn{1}{l|}{\multirow{2}{*}{F1}}&\multicolumn{1}{l|}{\multirow{2}{*}{GM}}&\multicolumn{2}{c|}{Risky}&\multicolumn{1}{p{.2cm}|}{Safe}
     &\multicolumn{1}{l|}{\multirow{2}{*}{F1}}&\multicolumn{1}{l|}{\multirow{2}{*}{GM}}&\multicolumn{2}{c|}{Risky}&\multicolumn{1}{p{.2cm}|}{Safe}
     &\multicolumn{1}{l|}{\multirow{2}{*}{F1}}&\multicolumn{1}{l|}{\multirow{2}{*}{GM}}&\multicolumn{2}{c|}{Risky}&\multicolumn{1}{p{.2cm}|}{Safe}
     &\multicolumn{1}{l|}{\multirow{2}{*}{F1}}&\multicolumn{1}{l|}{\multirow{2}{*}{GM}}&\multicolumn{2}{c|}{Risky}&\multicolumn{1}{p{.2cm}|}{Safe}\\
    	&\multicolumn{1}{p{.2cm}|}{}&\multicolumn{1}{p{.2cm}|}{}&Pre&\multicolumn{1}{p{.2cm}|}{Rec}&\multicolumn{1}{p{.2cm}|}{Rec}
    	&\multicolumn{1}{p{.2cm}|}{}&\multicolumn{1}{p{.2cm}|}{}&Pre&\multicolumn{1}{p{.2cm}|}{Rec}&\multicolumn{1}{p{.2cm}|}{Rec}
    	&\multicolumn{1}{p{.2cm}|}{}&\multicolumn{1}{p{.2cm}|}{}&Pre&\multicolumn{1}{p{.2cm}|}{Rec}&\multicolumn{1}{p{.2cm}|}{Rec}
    	&\multicolumn{1}{p{.2cm}|}{}&\multicolumn{1}{p{.2cm}|}{}&Pre&\multicolumn{1}{p{.2cm}|}{Rec}&\multicolumn{1}{p{.2cm}|}{Rec}\\ 
    	\hline
    	Real Pred
    	&25~+&38~+&74&15&98
        &35~+&48~/&66&24&95
        &59~/&\textcolor{cyan}{\textbf{76}}~+&45&88&66
        &42~+&\textcolor{cyan}{\textbf{71}}~/&29&74&69\\
        RFS MLP&
        \textcolor{cyan}{\textbf{57}}~/&\textcolor{cyan}{\textbf{70}}~/&51&65&76&
        48~/&54~/&36&74&40&
        59~/&67~/&47&78&58&
        47~+&66~+&38&61&72\\
        \hline
        LL noRB 
        &52~+ &66~+&44&64&69
        &50~/&\textcolor{cyan}{\textbf{64}}~/&40&67&62
        &\textcolor{cyan}{\textbf{60}}~/&69~/&50&75&64
        &45~+&66~+&34&69&64\\
        LL
        &  56   &  69   &  49   &  64   &  75
        &  \textcolor{cyan}{\textbf{52}}   &  59   &  39   &  77   &  46
        &  59   &  66   &  45   &  85   &  52
        &  \textcolor{cyan}{\textbf{52}}   &  \textcolor{cyan}{\textbf{71}}   &  42   &  67   &  74\\
        \hline
    \end{tabularx}
\end{table*}

%% file: content/results.tex
\input{tables/RQ1_hyper}

\section{RQ1: \RQone}

{\noindent}\textit{Motivation.} 
In this question, we empirically compare the performance of the LL setup with RFS approaches previously used by our industrial partner and with an NN-based RFS approach.
We thus address the two following subquestions:
\begin{itemize}
   \item RQ1.1: \RQoneone
   \item RQ1.2: \RQonetwo
\end{itemize}

{\noindent}\textit{Approach.} 
For each dataset, we used the MLP architecture and LL setup hyper-parameters with the best median F1-score performance out of 5 runs on the validation sets. For those best setups, Table~\ref{CL_hyper_res}, row ``LL'', shows the average prediction results out of 5 runs on the test sets, 
which we will compare with the RFS setup baselines. 
The tuned LL setup hyper-parameters are reported in Table~\ref{hyper_val}.
In RQ1.1, we use the Weekly XGB and Heuri XGB baselines for brown build prediction, and Real Pred for risk prediction. In RQ1.2, we use the RFS MLP baseline for both use cases. The latter baseline uses the same MLP model and scheduling approach as the LL setup, yet using different training and data selection approaches, as mentioned in Table~\ref{tab:trainingsetup} (RFS MLP). 
To assess the significance of the performance differences, we use, for each project, a Kruskal-Wallis statistical test ($\alpha=0.05$) to compare the F1-score and G-mean of the different runs of the LL setup with the other setups.

\vspace{2mm}{\noindent}\textit{Results for RQ1.1:} \textit{\RQoneone}

\textbf{The LL setup has an F1-score significantly higher (with large effect) than an RFS setup in 6 projects (+9-31\%), while performance is similar in the 2 other projects 
(+0-4\%).}
For brown build prediction, when comparing the LL performance with the weekly retraining of an XGBoost model (``Weekly XGB" in Table~\ref{CL_hyper_res}), the F1-score of the LL setup is significantly higher (+9-31\%) for all projects but Brown\_2.
Brown\_2's F1-score of 71\% still shows a small +4\% increase. 
The comparison to heuristics-based retraining (row ``Heuri XGB" line in Table~\ref{CL_hyper_res}) is very similar, again with statistical improvements for all projects but Brown\_2.

For JIT risk prediction, we compare the LL setup with our industrial partner's historical model performance (``Real Pred'' in Table~\ref{CL_hyper_res}). 
The F1-score of the LL setup is significantly higher with large effect size (+10-31\%) for three projects and identical for Risk\_3. 
While Risk\_1 and 2 show a decrease in precision (-25\% and -27\%) compensated by an increase in recall (+49\% and +43\%), Risk\_3 and 4 show the opposite, with 
 respectively no improvement and improvement (+13\%) in precision and a reduction in recall (-3\% and -7\%).

We can see that, overall, the observations with G-mean are the same as those of the F1-score for all the projects but two, with Risk\_3's real prediction performing significantly better than the LL setup and Risk\_4's real prediction behaving similarly. Note that we tuned the hyper-parameters based on the F1-score, which might explain these observations.

\noindent\doublebox{%
    \parbox{.46\textwidth}{%
        \textbf{RQ1.1: The LL setup performs at least similarly to the state-of-the-practice baselines for all datasets and significantly better (large effect) in 6/8 of the cases.}
    }%
}

\vspace{2mm}{\noindent}\textit{Results for RQ1.2:} \textit{\RQonetwo}

\textbf{3/8 datasets show a significant improvement in F1-score (+5-8\%) for the LL setup compared to an RFS MLP, 4/8 show no significant difference (-1\% to +8\%), and 1/8 shows a significant decrease (-5\%).}
In Table~\ref{CL_hyper_res}, we observe that using an LL setup significantly improves the F1-score in comparison to RFS MLP for Brown\_1 (+8\%), Brown\_OSS (+5\%), and Risk\_4 (+5\%). In contrast, Brown\_3 shows a significant decrease in F1-score (-5\% significant). In the other cases, LL performed at least as well as RFS MLP.
Regarding the G-mean, the same observations can be made apart from Brown\_1 not having a significant improvement, still achieving a +8\% increase. 

In the case of Brown\_3 (where LL performed significantly lower with large effect size), 
we noticed a decrease in precision (-5\%) and recall (-4\%) for the LL setup. This shows that in this case, seeing all of the past data actually improved the performance, which we suspect is due to the highly unbalanced nature of that dataset ($\beta$ of 5\%), causing LL to overtrain on the few brown jobs, while RFS has access to all historical jobs.
On the other hand, Brown\_OSS show huge improvement (+47\%) in F1-score between the RFS and LL, which indicates that the model trained on mostly recent datapoints improved the performance and forgetting past datapoints improves the model.
Two of the four projects with equal performance have a heavily cyclic development process of 3-month sprints; future empirical analysis should investigate the potential link between these process characteristics and the model results.

Overall, RQ1(.2) shows that the LL setup is able to match and in most studied cases even outperform the performance of previously used approaches.
Extending the LL setup with an online bagging approach~\cite{cabral2019class, cabral2022towards} would likely improve the performance even further~\cite{goodfellow2013empirical}, though also increasing the computational effort needed to train the multiple models in the bag.

\noindent\doublebox{%
    \parbox{.46\textwidth}{%
        \textbf{RQ1.2: The LL setup performs at least similarly to NN-based RFS in 7/8 cases and is outperformed in 1 case.
        Hence, LL can bring value in terms of model performance, but is not a panacea.
        }
    }%
}

\section{RQ2: \RQtwo}

{\noindent}\textit{Motivation.} LL setups are often at risk of showing symptoms of catastrophic forgetting, gradually decreasing the performance of LL retraining. To reduce this risk, we adopt a replay buffer (memory-based approach). In this question, we aim at evaluating how impactful this approach is on LL.

{\noindent}\textit{Approach.} We compare the LL setup with(out) replay buffer, corresponding respectively to rows ``LL"/``LL noRB" in Table~\ref{CL_hyper_res}.

{\noindent}\textit{Results.} \textbf{Using a replay buffer significantly improved the F1-score for 3 projects (+4\% to +26\%), without significant change in performance for the other projects (-1\% to +6\%).}
In Table~\ref{CL_hyper_res}, we observe that using a replay buffer significantly improves the F1-score (with large effect size) for Brown\_OSS (+26\%), Risk\_1 (+4\%), and Risk\_4 (+7\%). While Risk\_3 sees an F1-score decrease by 1\%, the difference in performance is not significant. 
The G-mean observations are similar to those of the F1-score.

We thus observe less improvement than Gao et al., whose replay approach improved results in all cases~\cite{gao2023keeping}. 
However, their NLP models take as input text, while our approaches respectively take pre-embedded input (brown builds) or feature-based inputs (risk prediction), both of which might capture less change over time than text does due to the abstraction of those inputs

\noindent\doublebox{%
    \parbox{.46\textwidth}{%
        \textbf{RQ2: The use of a memory-based approach (replay buffer) to reduce catastrophic forgetting significantly improves the performance for 3 out of 8 projects, while maintaining performance for the others.}
    }%
}

\section{RQ3: \RQthree}

{\noindent}\textit{Motivation.} 
The computational effort of model updates in an LL setup is lower than RFS since less data is considered.
However, depending on the retraining frequency in both setups, the combined computational effort across all model updates might outweigh the single model retraining computation of RFS. 
This might make LL less tempting to adopt in the industry, especially in cases where LL did not outperform the RFS setups.
Hence, in this RQ, we evaluate how often updates are necessary as well as approximate the actual computational effort of update/model training.

\vspace{2mm}{\noindent}\textit{Approach.} We model the overall computational effort for both retraining setups in terms of the life expectation of models and the complexity of the retraining algorithms~\cite{wilf2002algorithms}. First, the life expectation (LifeExp) of a model corresponds to the time between successive model updates based on the maximum timestamps of training and validation data points for each training.
For each project and a given retraining setup, Table~\ref{time_group} shows the median life expectation across all retraining attempts. 
Secondly, as done by Thuijsman et al.~\cite{thuijsman2019computational}, we calculate the algorithms' complexity in terms of the size of the training set $TrainS$.
As shown in Algorithm~\ref{algo_nn}, the update/training time of our MLP models is linearly dependent on the size of the training set, as is the complexity of the XGBoost training algorithm~\cite{chen2016xgboost}. In other words, we can compare the computational effort of RFS and LL retraining setups by comparing the training data sizes. 

To combine both measures into one measure of computational effort, we define the effort coefficient $coef$=$\frac{trainS}{lifeExp}$ and the relative coefficient $coef_{rel}$=$\frac{coef}{coef[LL]}$, relative to the coefficient of the LL setup. The higher the $coef_{rel}$ for a given training setup, the more effort it requires compared to LL. In particular, in RQ2, we compare the ``RFS MLP'' and ``LL'' setups, as they both train the same type of MLP and just differ in the update computational effort.

{\noindent}\textit{Results.} \textbf{The median life expectation of models with an LL setup is between 1.1-8.75 times shorter than the RFS MLP baseline for 7 of the studied projects and is 2.3 times longer for Brown\_OSS.}
Table~\ref{time_group} shows how, in the case of brown build prediction, the median life expectation of LL models for Brown\_1, 2 and 3 is 2.8-8.75 times shorter than for Heuri XGB, needing thus more updates for similar to significantly better model performance (cf. RQ1). On the other hand, for Brown\_OSS, the median life expectation of models is 2.3 times longer than for both XGB-based baselines because that project has much slower brown build activity, if we refer to the total age of the project and the amount of brown jobs from Table~\ref{info_proj}. Since less frequent brown jobs are encountered, training more often (with smaller $GS$) would most likely overfit on the lower amount of brown jobs present in the training set.

For JIT risk prediction, the LL model needs updates every .6-1 month as a median, whereas the actual RFS is done every 1.5-2 months by our industrial partner (Real Pred), yielding thus a median life expectation 1.1-2.2 times shorter for the LL setup.

\textbf{The LL updates use about 3.3-13.7 times less data during updates than the RFS approach.}
Similar to Olewicki et al.~\cite{olewicki2022towards}, we found that the RFS approach for brown build detection requires an optimal training dataset size of 30 weeks for all studied projects. To put this in context, we express this in terms of the median $GS$ (in terms of number of commits) of data groups used by LL in Table~\ref{datasets_size} (row ``RFS XGB (med)"). 
The LL initialization step (which is equivalent to RFS) uses a training set of $ITW$ time groups (see Table~\ref{hyper_val}), with each group containing $GS$ data points. We consider the resulting $\sim 10$-$15~GS$ data points as a lower bound for the dataset size in an RFS MLP setup.
Later model updates use a training set composed of both new data points since the last update and a replay buffer, amounting to $GS + \frac{RBSSize}{100}\times RBWin \times GS$ data points. 

When comparing in Table~\ref{datasets_size} the training set sizes of the LL updates to RFS XGB for brown build detection 
, the dataset of LL updates is 3.3-13.7x smaller (\eg $24.6/1.8$=$13.7$ for Brown\_1). Then, when comparing LL updates with LL initialization (row ``LL Speed up''), this time on all 8 projects, we find that the dataset of LL updates is 5-8.3x smaller than the LL initialization.

\textbf{Figure~\ref{fig:coef_rel} shows that the relative effort coefficient $coef_{rel}$ for all the setups is between 2-40}, demonstrating the overall lower computational effort of the LL setup.
With alternative state-of-the-art approaches, such as online bagging~\cite{cabral2019class}, the computation effort would be multiplied by the number of models in the bag. There is thus a trade-off to be made between improving the performance with bagging, while increasing the computational retraining effort.

\textbf{The actual retraining time speedup matches the theoretical one for Risk\_2\&3 datasets, and is lower than expected for the other datasets, although still showing at least a 3.6-fold speedup compared to RFS updates.}
Table~\ref{datasets_size} shows 
the actual mean computation time for initialization and updates of the MLP model in an LL setup for the Risk datasets. The LL Speedup ratio ($\frac{initializat.}{updates}$) is given to compare LL to RFS computational effort, as each RFS retraining takes a similar time as the LL initialization time. 
In half of the projects, the speedup in terms of time matches the theoretically expected speedup (linear dependency on data file size), while the LL time efficiency, though lower than expected, is still at least 3.6x as high for the others. The reason for the lower speedup is the non-optimized memory management in the Python libraries causing extra time consumption in the training steps, independent of the size of the data.

\noindent\doublebox{%
    \parbox{.46\textwidth}{%
        \textbf{
        RQ3: The effort of maintaining an LL setup is 2-40x less computationally expensive than an RFS setup. 
        }
    }%
}

\input{tables/lifeexp}

\input{tables/datasize}

\input{figures/rel_coef}

%% file: tables/RQ1_hyper.tex
\begin{table}[t]
	\centering
	\caption{Optimal LL hyperparameter values used in our study.}
	\label{hyper_val}
	\begin{tabular}{|p{2.5cm}|p{.3cm}|p{.3cm}|p{.3cm}|p{.4cm}|p{.3cm}|p{.3cm}|p{.4cm}|p{.3cm}|}
		\hline \multirow{3}{*}{Name}& \multicolumn{8}{c|}{Project's tuned hyperparameters' values}\\\cline{2-9}
	    &\multicolumn{4}{c|}{Brown\_}&\multicolumn{4}{c|}{Risk\_}\\
		\cline{2-9}
		&1&2&3&OSS&1&2&3&4\\\hline

		Group size $GS$&50	&50	&50	&100
		&500&500&1000&500\\
		Initial train window $ITW$&15	&10	&10	&10
		&15&15&15&15\\
		Validation window $VWin$&15		&10		&10		&10
		&8&15&8&8\\
		Replay Buffer sample size [\%$GS$] $RBSize$	&10	&5	&5	&20
		&10&10&10&10\\
		Replay buffer window $RBWin$	&8	&10	&5	&10
		&8&15&8&8\\
		\hline
	\end{tabular}
\end{table}

%% file: tables/lifeexp.tex
\begin{table}[t]
    \centering
    \caption{Median life expectation of models in \#days.}
    \label{time_group}
    \begin{tabularx}{.47\textwidth}{|l|X|X|X|l|}
    	\hline
    	Life~Exp~[day]&Brown\_1&Brown\_2&Brown\_3&Brown\_OSS\\ \hline
    	Weekly XGB & 7&7&7&7\\
    	Heuri XGB&42&35&14&7\\
    	LL (med.)&9&4&5&18\\
    	LL (avg.)&12$\pm$3&12$\pm$9&6$\pm$2&85$\pm$99\\
    	\hline
    	\hline
    	Life~Exp~[day]&Risk\_1&Risk\_2&Risk\_3&Risk\_4\\ \hline
    	Real Pred &$49\pm$7&$49\pm$7&$49\pm$7&$49\pm$7\\ 
    	LL (med.)&32&22&35&36\\ 
    	LL (avg.)&40$\pm$11&32$\pm$6&43$\pm$6&32$\pm$13\\ 
        \hline
    \end{tabularx}
\end{table}

%% file: tables/datasize.tex
\begin{table}[t]
    \centering
    \caption{Size of training sets for model training/updates in terms of $GS$ for all datasets, and average model computation time across 5 runs for the Risk datasets (training with 50 epochs and 10 sample batches).}
    \label{datasets_size}
    \begin{tabularx}{.47\textwidth}{|l|X|X|X|c|}
    	\hline
    Dataset size &Brown\_1&Brown\_2&Brown\_3&Brown\_OSS\\ \hline
    	RFS XGB (med)& 24.6$GS$&11.4$GS$&10.5$GS$&6.6$GS$\\
    	LL initializat.&15$GS$&10$GS$&10$GS$&10$GS$\\
    	LL updates&1.8$GS$&1.5$GS$&1.25$GS$&2$GS$\\
        LL Speedup&8.3$\times$&6.7$\times$&8.$\times$&5.$\times$\\
    	\hline
    	\hline
    	Dataset size &Risk\_1&Risk\_2&Risk\_3&Risk\_OSS\\ \hline
    	LL initializat.&15$GS$&15$GS$&15$GS$&15$GS$\\
    	LL updates&1.8$GS$&2.5$GS$&1.8$GS$&1.8$GS$\\ 
        LL Speedup&8.3$\times$&6.$\times$&8.3$\times$&8.3$\times$\\
            \hline
        Avg. time &Risk\_1&Risk\_2&Risk\_3&Risk\_OSS\\ 
     \hline
        LL initializat.&38.6 s&121.4 s&91.9 s&32.7 s\\
        LL updates&10.4 s&12.7 s&11.4 s&9. s\\ 
        LL Speedup&3.7$\times$&9.6$\times$&8.1$\times$&3.6$\times$\\
        \hline
    \end{tabularx}
\end{table}

%% file: figures/rel_coef.tex
\begin{figure}
     \centering
    \includegraphics[width=.45\textwidth]{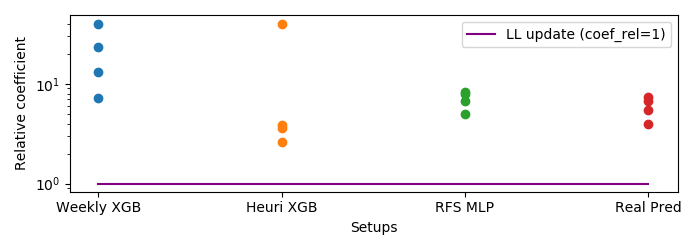}
    \caption{Relative coefficient $coef_{rel}$ of training setups compared to the LL setup updates.}
    \label{fig:coef_rel}
\end{figure}

%% file: content/discussion.tex
While software analytics tools are being adopted in industry to ease the work for developers, the design and deployment of such tools do not take into account the evolving nature of software projects, instead relying on stationary ML models (and RFS).
Through our RQs, we observe that the performance of an LL setup can match, if not improve, that of an RFS setup, while considerably reducing the computational effort of updating the model by a coefficient $coef\_rel$ of 2-40.
The speedup of individual model updates of the LL setup compared to RFS reduces energy consumption but also allows more frequent updates to keep the models up-to-date, addressing drift.

One aspect of software analytics models that was ignored until now is the human interpretation of said models, as an updated model might convey a contradictory message to developers compared to prior versions, for example when the age of a file at one time step would correlate with higher risk, yet at the next step would correlate with lower risk.

Hence, to analyze the relation of LL to the potential cognitive effort of changing model interpretation, 
we evaluate how the feature importance of models in LL and RFS setups evolves over time using the \texttt{captum}~\cite{captum} Python package. 
Feature importance is a metric that shows the strength of the learned correlation of a feature with the prediction~\cite{captum}.
Figure~\ref{fig:feat_importance_1_2} shows how it evolves in the LL and RFS setups for the 10 features with the highest sum of absolute average and variance of feature importance. 
To better observe how feature importance can change from one extreme to another, we discretize the feature importance values. Positive importance values are colored in red ($>.1\times Max$), negative importance in blue ($<-.1\times Max$) and intermediate (``Mid'') in gray, where $Max$ is the maximum absolute value of feature importance at a given time step for a given project.

From Figure~\ref{fig:feat_importance_1_2}, we observe that sometimes the LL setup's feature importance varies less over time than the RFS setup's (e.g., Reviewers relative experience), while at other times we see the opposite (e.g., Reviewers experience). In order to quantify this observation, Table~\ref{tab:feat_perc} provides the percentage of features that have a larger feature importance variance, calculated as $\frac{1}{|F|}\sum_{i\in F} (f_i - mean(F))^2$, with $F$ the vector of feature importance across time steps. 
The LL setup has +6\% to +58\% more features with higher variation than RFS for Risk\_1\&2\&3, while 40\% less for Risk\_4.
However, when looking at the discretized feature importance values (using +.1, 0. and -.1 values $f_i$ for Pos, Mid, Neg, respectively), only Risk\_1 has a higher percentage of features varying more in the LL setup (+6\%), while the other projects show the opposite (from +24\% to +52\% for RFS).
This confirms our observations of LL vs. RFS in Figure~\ref{fig:feat_importance_1_2}.

\input{figures/disc_feat_importance}

In other words, either setup could actually lead to more stable or flexible models in terms of feature importance. The LL setup gains stability because all updates incrementally derive from one initial model, while it gains flexibility from seeing a smaller amount of relatively recent data (new data and the replay buffer). On the other hand, the RFS setup gains flexibility from having a new model retrained from scratch at each step, but gains in stability from training on the complete dataset available at each training point.

The fact that the discretized analysis of feature importance variance shows more stability for LL relative to RFS than the non-discretized analysis indicates that many of the fluctuations in feature importance for LL occur within discretized categories (Pos, Neg or Mid), corresponding to minor changes in model interpretation, while major changes crossing borders between Pos/Mid/Neg are less common than for RFS. Hence, models in LL setups seem to have a relatively good continuity.
Then, whether this continuity is ``correct'' or if the LL setup misses important feature importance changes, the fact that the LL setup often showed better F1-score and/or G-mean performance than the RFS setup hints at the former.

%% file: figures/disc_feat_importance.tex
\begin{figure}[t]
     \centering
     \begin{subfigure}[b]{0.45\textwidth}
         \centering
         \includegraphics[width=\textwidth]{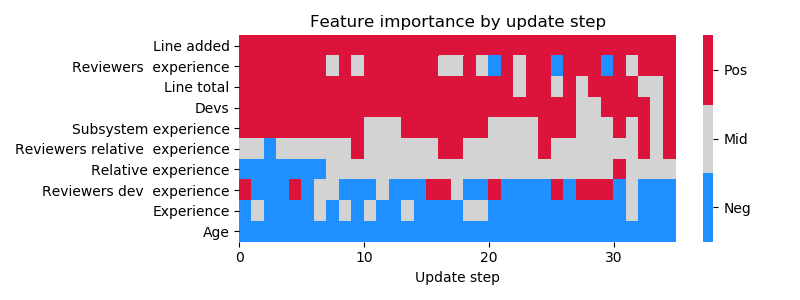}
         \caption{LL setup for Risk\_1}
         \label{fig:feat_importance_ll_1}
    \end{subfigure}
     \hfill
     \begin{subfigure}[b]{0.45\textwidth}
         \centering
         \includegraphics[width=\textwidth]{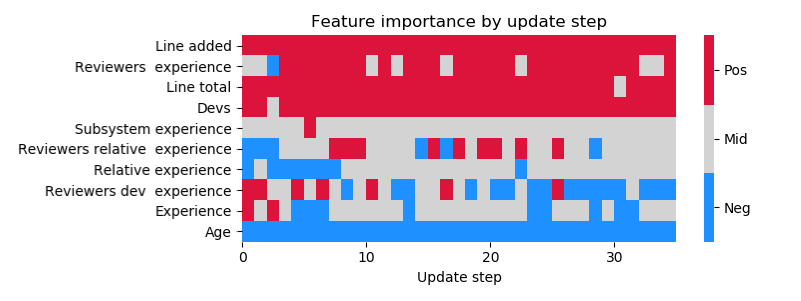}
         \caption{RFS setup for Risk\_1}
         \label{fig:feat_importance_micro_1}
     \end{subfigure}
     
     \caption{Discretized feature importance evolution for Risk\_1. \label{fig:feat_importance_1_2}}
\end{figure}

%% file: content/threats.tex
\input{tables/disc_feat}
\textbf{External validity.}
We evaluate LL setups in an industrial context regarding performance and required maintenance effort, for two use cases brown build detection and JIT risk prediction, as exemplars of software analytics classification models used in industry. 
Overall, the LL approach was applied to eight projects, among which seven from our industry partner, Ubisoft. 
More projects and analytics tools should be studied to validate the generalization of our results to the wider domain of software analytics.

\textbf{Construct validity.}
With the current LL setup, the feature selection step of the brown build detection tool is only done at the initialization step.  
This means that if new features become relevant over time, those will not be considered by the model. 
Future research could improve on this by using, for instance, flexibly-sized MLP models that would allow LL updates while adding new nodes when needed (and thus adapt to changing inputs)~\cite{kirkpatrick2017overcoming}. 
This is not an issue for JIT risk prediction, which has a fixed set of features.

We evaluated the computational effort of LL computed to RFS in RQ3, considering single MLP model architectures. It can be noted that in the case of using bagging~\cite{cabral2022towards}, the computational effort would be multiplied by the size of the bag for both setups.

We used the time ordering of commits for data splitting, but our analyses did not take into account the delay between a prediction and the availability of its true JIT risk label. 
However, since we update the model every $GS$ data points, many new data points would have had time to be correctly labeled in the meantime. 
Also, the JIT models yielded similar improvements compared to the baseline as our Brown build detection models, for which no delay in labeling occurs (since humans evaluated predicted risk cases immediately). This suggests that the impact of the delay issue might be limited.

Our approach was optimized regarding the F1-score metric, which considers precision and recall as equally important. Though the precision or recall could be improved by changing the threshold of decision, chosen to compute the F1-score. We used a default 50\% threshold. By increasing the threshold, the precision would increase. As precision and recall are connected, this increase would also lead to a decrease in recall and vice-versa.

%% file: tables/disc_feat.tex
\begin{table}[t]
    \centering
    \caption{\%Features with higher variance of importance for RFS or LL compared to the other for the Risk use case. The variance is computed on the feature importance value before discretization (first 3 rows) and after (last 3 rows).}
    \label{tab:feat_perc}
    \begin{tabularx}{.48\textwidth}{|l|X|X|X|X|}
    \hline
    
      [\%]&Risk\_1&Risk\_2&Risk\_3&Risk\_4\\
    \hline
    for RFS (not discretized) & 21&21&36&58\\
    for LL (not discretized) & 79&58&42&18\\
    Same (not discretized) & 0&21&21&24\\
    \hline
    for RFS (discretized) &30&48&58&36\\
    for LL (discretized)&36&9&6&12\\
    Same (discretized) &33&42&36&52\\
    \hline
    \end{tabularx}
 \end{table}

%% file: content/conclusion.tex
More and more ML-based software analytics tools are used in industry as a practical way of automating software engineering tasks.
However, in practice, as a model trained at the beginning of the development process might not be relevant later on due to drift, these ML models are regularly being discarded in lieu of new models retrained from scratch.
Ubisoft, our industrial collaborator, identified this issue for their software analytics tools. 
By using an LL setup, the same model could be updated incrementally based on the previously pre-trained model and only a limited historical dataset necessary to avoid catastrophic forgetting.
By evaluating the LL setup on eight projects, seven industrial and one open-source, across two software analytics use cases, we observed that LL performs at least as well as RFS in most of the cases and often outperforms it in practice.
Also, in contrast to recent work, we find that a replay buffer is not always necessary to avoid catastrophic forgetting in our use cases, though improving performance significantly in 3/8 cases.

More importantly, we observe an important gain with LL setups as they are computationally less expensive to maintain.
We found that the LL setup leads to a considerable speed-up of the training process as updates use 3.3-13.7 times less data than RFS setups. This enables LL retraining to be performed more frequently without increasing the total computational effort of retraining, effectively making LL both feasible and valuable in practice.

Finally, we observe that LL setups have more stable predictions as the models are updated on top of past models, having a better continuity than RFS regarding feature importance. This reduces the cognitive effort of changing model interpretation for users.

